\documentclass[]{spie}  

\usepackage{amsmath,amsfonts,amssymb}
\usepackage{graphicx,aas_macros}
\usepackage[colorlinks=true, allcolors=blue]{hyperref}

\title{The science calibration challenges of next generation highly multiplexed optical spectroscopy: the case of the Maunakea Spectroscopic Explorer}

\author[a,b]{Alan W. McConnachie}
\author[a]{Nicolas Flagey}
\author[c]{Pat Hall}
\author[d]{Will Saunders}
\author[a]{Kei Szeto}
\author[a]{Alexis Hill}
\author[e]{Shan Mignot}
\affil[a]{The Maunakea Spectroscopic Explorer Project Office, 65-1238 Mamalahoa Hwy Kamuela HI 96743 USA}
\affil[b]{NRC Herzberg, Dominion Astrophysical Observatory, 5071 West Saanich Road, Victoria,
British Columbia, Canada}
\affil[c]{Department of Physics and Astronomy, York University, Toronto, ON M3J 1P3, Canada}
\affil[d]{Australian Astronomical Observatory, PO Box 915, North Ryde, NSW 1670, Australia}
\affil[e]{GEPI, Observatoire de Paris, PSL Research University, CNRS, Univ. Paris Diderot,
Sorbonne Paris Cit{\'e}, Place Jules Janssen, 92195 Meudon, FRANCE}

\authorinfo{Further author information: (Send correspondence to A.W.M.)\\A.W.M.: E-mail: mcconnachie@mse.cfht.hawaii.edu, Telephone: 1-808-885-3188}

\begin{document} 
\maketitle

\begin{abstract}
MSE is an 11.25m telescope with a 1.5 sq.deg. field of view. It can simultaneously obtain 3249 spectra at $R=3000$ from $360-1800$nm, and 1083 spectra at $R=40000$ in the optical. The large
field of view, large number of targets, as well as the use of more than
4000 optical fibres to transport the light from the focal plane to the
spectrographs, means that precise and accurate science calibration is
difficult but essential to obtaining the science goals. As a large
aperture telescope focusing on the faint Universe, precision sky
subtraction and spectrophotometry are especially important. Here, we
discuss the science calibration requirements, and the adopted
calibration strategy, including operational features and hardware, that will enable the successful scientific
exploitation of the vast MSE dataset.
\end{abstract}


\section{INTRODUCTION}
\label{sec:intro}  

The Maunakea Spectroscopic Explorer (MSE) is the only dedicated,
optical and near-infrared, large aperture ($> 10$\,m), multi-object
spectroscopic facility being designed for first light in the
mid-2020s. It is a re-purposing of the Canada-France-Hawaii Telescope,
within an expanded international partnership and upgraded to a larger
aperture.

MSE will obtain more than 4000 spectra per observation. Specifically,
photons will be collected by the 11.25m aperture M1, reflected to the
prime focus where a Wide Field Corrector will provide a 1.5 square
degree field of view. An Atmospheric Dispersion Corrector will correct
for some effects from the atmosphere, and at the focal plane more than
4000 fiber positioners will move fibers to the expected locations of
the astronomical targets, including science objects, calibration
sources, and sky positions. Those photons within the entrance aperture
of the fibers will pass down tens of meters of fiberoptic cable to one
of two different suites of spectrographs. There, the photons will pass
through numerous optics, including dispersive elements, before being
registered on the CCD or H4RG detectors.

Astronomers using MSE will only obtain an improved understanding of
the wonders of the Universe by successfully relating the counts
on detectors (the end result of the journey of the photons through the
atmosphere and the MSE system) to the physical properties of the
astrophysical objects that the photons left some large number of years
ago.  Calibration is therefore central to all aspects of MSE. By its
nature, accurate and precise calibration is challenging, and it is made doubly so for MSE which has
as its driving goals the study of exceptionally faint astrophysical
sources, including spectrophotometric studies and time-domain (long
baseline) observations.

At the previous SPIE Astronomical Telescopes and Instrumentation
meeting, the status and progress of the project were detailed in
Ref.~\citenum{murowinski2016} while an overview of the project design
was given in Ref.~\citenum{szeto2016} and the science based
requirements were explained in Ref.~\citenum{mcconnachie2016}. An
update of the project at the end of conceptual design phase is
presented this year in Ref.~\citenum{szeto2018a} with a review of the
instrumentation suite in Ref.~\citenum{szeto2018b}. Other papers
related to MSE are focusing on: the summit facility upgrade
(Ref.~\citenum{bauman2016, bauman2018}), the telescope optical designs
for MSE (Ref.~\citenum{saunders2016}), the telescope structure design
(Ref.~\citenum{murga2018}), the design for the high-resolution
(Ref.~\citenum{zhang2016, zhang2018}) and the low/moderate-resolution
spectrograph (Ref.~\citenum{caillier2018}, the top end assembly
(Ref.~\citenum{mignot2018, hill2018b}), the fiber bundle system
(Ref.~\citenum{venn2018, erickson2018}), the fiber positioners system
(Ref.~\citenum{smedley2018}), the systems budgets architecture and
development (Ref.~\citenum{mignot2016, hill2018}), the observatory
software (Ref.~\citenum{vermeulen2016}), the spectral calibration
(Ref.~\citenum{flagey2016a, mcconnachie2018a}), the throughput
optimization (Ref.~\citenum{flagey2016b, mcconnachie2018b}), the
observing efficiency (Ref.~\citenum{flagey2018b}), and the overall
operations of the facility (Ref.~\citenum{flagey2018a}).

This paper is structured as follows. Section 2 describes some of the key
science capabilities of MSE and their consequences for the
calibration requirements. Section 3 outlines the technical
considerations for sky subtraction and spectrophotometry , both in a general sense and specific to MSE
observations. Operational considerations
that affect calibration strategies are described in Section 4. In
Section 5, we present an overview of the baseline calibration
procedures for MSE, and in Section 6 we describe the initial concept
for the science calibration hardware. Section 7 discusses future
development plans and summarises.

\section{SCIENCE CAPABILITIES AND CONSEQUENCES FOR CALIBRATION}

   \begin{figure} 
   \begin{center}
   \includegraphics[height=8cm]{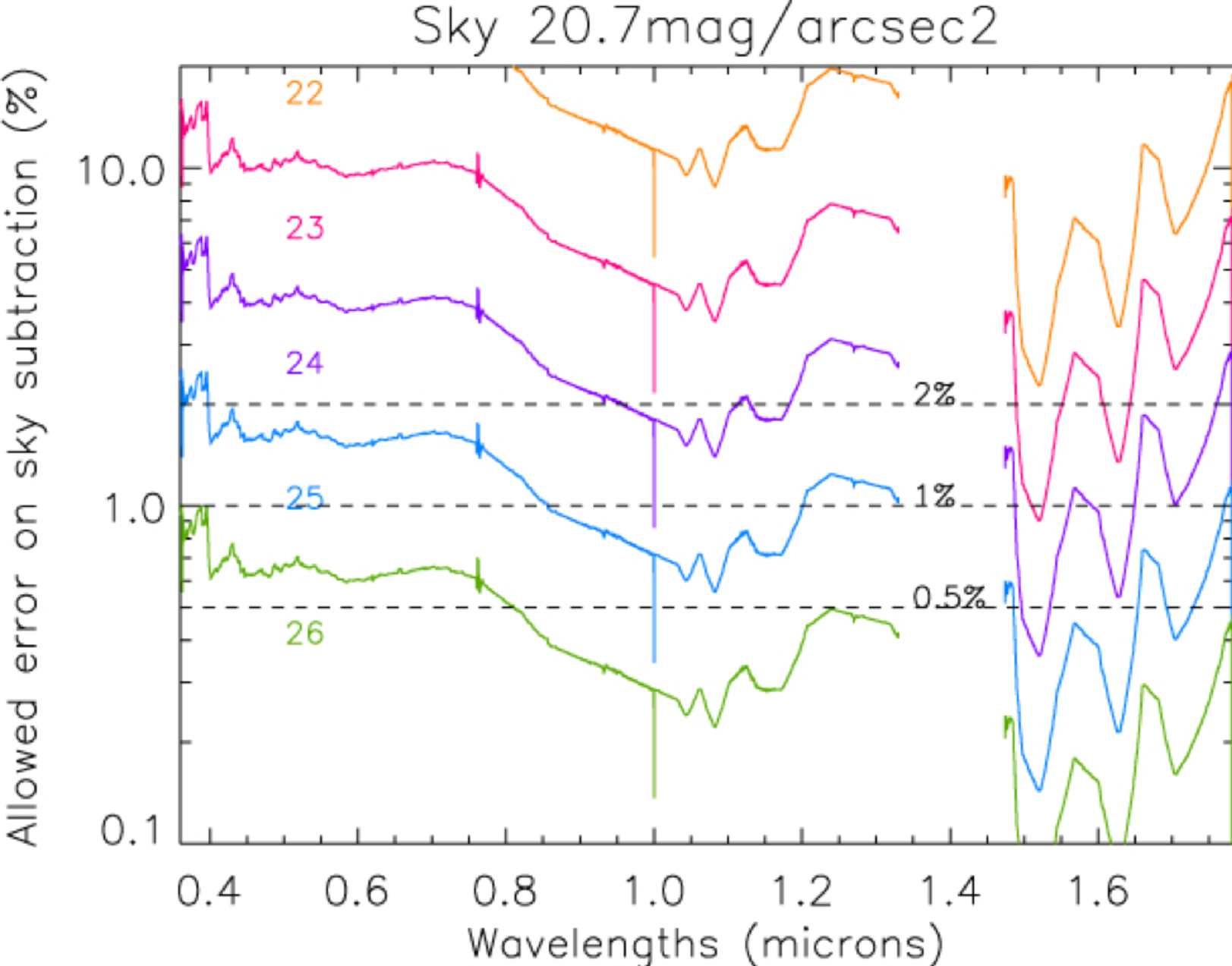}
   \end{center}
   \caption[example] 
   { \label{fig:sky} 
For dark sky conditions, each line shows the required sky subtraction
accuracy as a function of wavelength for sources with different
(monochromatic) magnitudes, to ensure that a majority of the flux is
the subtracted spectrum is from the target object, rather than
residual sky flux. }
   \end{figure} 

As a large aperture, dedicated, spectroscopic
facility, MSE is designed to obtain spectra of faint targets that
cannot be observed with smaller facilities. A multitude of possible
science programs for MSE are outlined in the Detailed Science
Case\cite{mcconnachie2016b}. Extragalactic
targets are typically 24th magnitude or even fainter; it is
already expected that MSE will obtain spectra on galaxies fainter than
25th magnitude. Dark skies in Maunakea correspond to around
20.7mags/sq.arcsec in the V-band. Thus, for unresolved or partially
resolved sources, we will routinely be targeting galaxies that are
more than 20 or even 50 times fainter than the sky.

Figure~\ref{fig:sky} shows the challenge for accurate sky subtraction as a
function of wavelength. Here, we consider astronomical targets with a
range of (monchromatic) intrinsic magnitudes, as indicated by the text
in the figure. Given dark time sky conditions ($V =
20.7$mags/sq.arcsec), the diffferent lines correspond to the
accuracy on the sky subtraction required at any wavelength in order to ensure that the
subtracted spectrum is dominated by target flux, not residual sky
flux. In almost all science cases, it will be necessary to require a 
significantly higher fraction of science photons that this (at least
10 times more). For discussion purposes, however,  we can consider it 
an absolute lower limit to what might be useful for science. For example, a
target at 25th magnitude at 850nm requires better than 1\% sky
subtraction in order to ensure a majority of the photons in the
subtracted science spectrum are actually science photons, not residual sky
photons. Clearly, to push to 25th magnitude or better requires sky
subtraction accuracies much better than 1\% at all wavelengths, and
the requirements are even more stringent in the H-band.

It is not just sky subtraction that will be challenging. A driving science goal of MSE is
reverberation mapping of Active Galactic Nuclei. This necessitates
accurate ($<3\%$) relative spectrophotometry i.e., the integrated flux
is a given wavelength interval compared to that in a different
wavelength interval should be able to be measured to $<3$\%
accuracy. This necessitates extremely accurate
knowledge of the transmission function of the system as a function of
wavelength at all times. Enabling this specific goal for reverberation
mapping enables many other science goals, especially in relation
to the stellar population modeling of galaxies and measurements of
their star formation histories.

Reverberation mapping also requires repeat observations with multiple
cadences, spread over a period of many years. Indeed, time-domain
science is expected to feature heavily in the science program of MSE. Accurate calibration over timescales
of years, to enable observations taken at widely seperated times to be
directly compared to one another, is a challenging but scientifically
compelling capability for which MSE is being designed.

There are 6 high level MSE Science Requirements that explicitly focus
on calibration, specifically velocity accuracy, spectrophotometric
accuracy, and sky subtraction. The calibration requirements, like all
science requirements, have been derived from consideration of the
science described in the Detailed Science Case, and in particular the
Science Reference Observations. The SROs describe, in considerable
detail, specific and transformational science programs that are
uniquely possible with MSE. Science requirements are defined as the
core set of science capabilities that MSE must have in order to enable
the science described by the SROs.

The Science Requirements relating to calibration are the following:

\begin{itemize}
\item REQ-SRD-041/042/043 Velocities at low/moderate/high resolution:
  For any object with a known velocity, observed at multiple epochs by
  MSE with up to a 5 year cadence with a signal to noise ratio per
  resolution element of 5/5/30 at low/moderate/high spectral
  resolution, the contribution from MSE to the rms difference between
  the known velocity of the object and the measured velocity of the
  object shall be less than or equal to 20/10/0.1 km/s

\item REQ-SRD-044 Relative spectrophotometry: For a spectrophotometric
  standard star, observed in the low resolution mode at multiple
  epochs by MSE with up to a 5 year cadence with a signal to noise
  ratio per resolution element of 30, the rms variation in the ratio
  of fluxes measured in any two wavelength intervals shall be less
  than 3\% of the mean measured value.

\item REQ-SRD-045 Sky subtraction, continuum: In wavelength intervals
  free from airglow emission-line contamination and strong telluric
  absorption, MSE shall allow for removal of the sky flux with a
  root-mean-square error of less than 0.5\% of the sky flux, at all
  wavelengths (TBC).

\item REQ-SRD-046 Sky subtraction, emission lines: MSE shall achieve a
  sky subtraction accuracy for atmospheric airglow emission-lines such
  that the mean residual error for spectral pixels, within 1
  resolution element of known atmospheric emission-lines, is $<1.5$
  times (TBC) the Poisson limit indicated by the propagated variance
  spectrum for each resolution element.
\end{itemize}

In what follows, we develop a calibration procedure based on three principles:

\begin{enumerate}
\item Calibration exposures shall not introduce any significant
  sources of noise into the observations (either directly such as a
  spectral flat or indirectly such as through a model of the flux
  response across the focal plane)
\begin{itemize}
\item The number of counts in a calibration exposure at any wavelength
  that is combined with the science data should greatly exceed that of
  the typical counts in the data (targets+sky combined). 
\end{itemize}
\item Calibration exposures shall not add significant overheads to science observations
\begin{itemize}
\item Any (standard) nighttime calibration observations will be quick,
  and any necessary time-consuming calibration should ideally be done
  during the day. It is a high level science requirement (see next section) that the nighttime science observing efficiency (i.e., open shutter time on science targets, excluding weather) should exceed 80\%. 
\end{itemize}
\item Calibration exposures shall be obtained in a configuration and
  under circumstances that are as close as possible to that of the
  science observations
\begin{itemize}
\item To reproduce as closely as possible the system-wide behavior at
  the time of the science observation
\end{itemize}
\end{enumerate}

\section{TECHNICAL CONSIDERATIONS}

\subsection{ General considerations}
 
In order to provide high quality science calibration, it is essential
to understand the wavelength-dependent transmission of MSE (the
transfer function) and the behaviour of the point-spread function for
astronomical targets positioned anywhere in the sky, in addition to
the wavelength solution. Thus, for every target, we want to know the
throughput, wavelength solution (mapping detector pixels into
wavelength), and the behaviour of the point spread function, as a function of

\begin{itemize}
\item Wavelength ($\lambda$)
\item Telescope pointing [Azimuth (a), Altitude (A)]
\item Position in field (on focal plane) [$X,Y$]
\item Position of fiber in its patrol region [$x,y$] 
\item Time ($t$)
\item Environmental conditions (temperature, humidity, etc.)
\end{itemize}

In practice, this means understanding the performance of various MSE subsystems as a function of all
these variables. Relevant subsystems include (considering successive ``Product Breakdown Structure''
elements along the optical path):

\begin{enumerate}
\item MSE.ENCL: the enclosure
\item MSE.TEL.STR: the telescope structure
\item MSE.TEL.M1: the primary mirror
\item MSE.TEL.PFHS: the prime focus hexapod system
\item MSE.TEL.WFC/ADC: the wide field corrector and atmospheric dispersion corrector
\item MSE.TEL.InRo: the instrument rotator
\item MSE.SIP.PosS: the fibre positioner system
\item MSE.SIP.FiTS: the fibre transmission system 
\item MSE.SIP.LMR or MSE.SIP.HR: the spectrographs, either
  low/moderate (these two modes are realised in a single spectrograph
  system) or high resolution
\item MSE.SIP.SCal: the science
  calibration hardware, specifically the calibration lamps and
  associated infrastructure
\item MSE.SIP.PESA: the program execution software architecture,
  including the data reduction and processing software and pipelines
\end{enumerate}

Our focus in this contribution is the strategy, operational
considerations and hardware that will allow us to obtain the best
empirical measurements of the PSF, throughput and wavelength
solutions. We do not discuss the data processing techniques by which
the calibration and science data will be combined (``PESA''), and leave this as
the subject of a future contribution.

A major calibration concern is is with respect to the behavior of the
fibers, and related components (e.g., positioners), in particular our
ability to map measurements made in one fiber at one time (with a
particular pointing and at a particular position in the field, where
the system, especially the fibers, are experiencing a specific
flexure, twist, etc.), to any other fiber (including the same fiber)
with a different pointing and/or at a different position in the
field. Consider two extreme scenarios:

\begin{itemize}
\item In an idealized perfect case, each fiber would be identical and
  have identical characteristics regardless of the details of the
  observation. Thus, it would be trivial to relate observations taken
  in one fiber (say, a spectrophotometric calibration star) to
  calibration information for another fiber. It would also be trivial
  to apply the wavelength solution derived for a fiber at one time to
  the same fiber at any future point in its usage.

\item In an idealized terrible case, each fiber would be very
  different, and would behave very differently depending on the
  details of the observation. Worse, this behavior would be
  non-repeatable. In such a situation, calibration observations such
  as arcs and flats would ideally be taken simultaneously (although in
  practice they would be taken immediately after or before the science
  observation) in order to estimate the behavior of the fiber at the
  particular moment of the observation.
\end{itemize}

We expect MSE and most other fiber MOS systems to be somewhere between
these two extremes, and the requirements of the system should seek to
make MSE as close as possible to the idealized perfect case. However,
each fiber will be intrinsically slightly different. We expect that
the behavior of MSE as a system (and the fiber system in particular)
will vary as a function of telescope pointing and target position
within the field, but we expect/require that this behavior will be
repeatable. Given the need to calibrate MSE to a very high level of
precision, we are cognizant that these effects may force us to
consider the MSE system hardware as closer to “terrible” than
“perfect”, and allowance must be made to accommodate for this
possibility.

\subsection{Considerations for sky subtraction}

Sky subtraction with fiber-fed spectrographs is usually performed by
allocating some number of fibers to measure the sky background
spectrum in blank locations across the field of view.  These 1-D sky
spectra are then combined, scaled to match the expected sky spectrum
in each target fiber, and subtracted from the 1-D target spectra.

The sky background consists of a smooth continuum plus airglow
emission lines, mostly from OH and mostly at wavelengths $>0.6$
microns.  The emission lines can vary on timescales of minutes
(Section 2.1.2 of Ref.~\citenum{ellis2008}) and on spatial scales of
arcminutes\cite{rodrigues2012}.  It is therefore necessary to obtain
sky spectra at the same time as the target spectra.

The sky background entering the fibers will vary depending on the
fiber's position in the field of view.   The sky background recorded
for each fiber on the detector will also depend on the
wavelength-dependent throughput of the fiber + instrument + detector
system for that fiber.  A wavelength-dependent scaling will be applied
to each spectrum to bring them to a common background level and a
common throughput.  Sky subtraction is limited by the accuracy of such
scaling, which will be limited by the accuracy of our measurements of
fiber positioner characteristics, of radial plate scale changes, and of the
wavelength-dependent system throughput for each fiber.

To help measure the rapidly changing sky background, one of two
varieties of "beam-switching" is sometimes used, at the cost of
observing efficiency.

\begin{itemize}
\item In telescope-based beam-switching, the telescope is offset between
exposures so as to place each target fiber on a blank sky position.
By assigning half the fibers to targets in each exposure, half the
fibers measure sky in each exposure, and each target has blank sky
measurements through the same fiber (at somewhat different times),
removing the need for spatial interpolation of the sky spectrum across
the field of view (FOV).  The disadvantage of this approach is that
exposure times per field become twice as long.
\item In fiber-based beam-switching, each target has two fibers
  assigned to it.  One is placed on the target and the other is placed
  on a blank sky location as close by as possible (and certainly
  within 60 arcseconds).  The allocation of one sky fiber to each
  target removes the need for spatial interpolation of the sky
  spectrum across the FOV.  The fibers assigned to each location can
  be swapped between exposures, so that the target and sky fluxes are
  measured through both fibers, to help with scaling sky spectra
  before subtraction.  The disadvantage of this approach is that the
  number of fibers available per field for science targets is cut in
  half.
\end{itemize}

Beam-switching is not planned to be the default MSE operational mode.
Nonetheless, to test the improvement (if any) obtained by fiber-based
beam-switching, it is recommended that MSE have the capability to
assign sky and object fiber pairs which could be alternated between
observations.

We now consider some more detailed specifics relating to sky
subtraction. We require calibration data that can provide wavelength solutions
and line spread function information for all fibers during the
observation.  Thus, arc spectra and lamp
flats taken close to the science observation are useful. 

The night sky background flux is uniform over the input face of a
fiber.  Thus, the PSFs of spectrally unresolved sky lines (in both the
spatial and spectral directions) will be determined by the exit beam
from the fibers and the spectrograph optics.

The spatial component of the PSF on the detector will affect sky
subtraction in 1-D spectra because spectrograph distortions cause the
spatial and spectral directions in a given 2-D spectrum to deviate
from lying exactly along detector rows or columns.

To correctly estimate the sky spectrum in one fiber from sky spectra
measured in other fibers requires knowledge of both the relative
wavelength-dependent sensitivities of the fibers and how the PSF
changes across each detector, in each spectrograph, with temperature
and mechanical changes, and with fiber illumination and focal ratio
degradation (see appendix A1 of Ref.~\citenum{sharp2010}).

The PSF can in principle be affected by:

\begin{enumerate}
\item Fiber tilt: For MSE, tilting spine positioners - ``Sphinx'' - are being used after an extensive
trade-study comparing the anticipated performance of these compared to
phi-theta positioners. Tilted spine positioners have been used by
Subaru/FMOS (``Echidnas''), and are to be used in 4MOST (``Aesop'').

Tilted spines hold the fibers very gently in comparison to ph-theta
positioners, and no twisting of the fiber occurs. However, for most
configurations the light enters the fiber at an angle. Light entering
a tilted fiber has a different angular distribution relative to the
fiber normal than light entering a face-on fiber.  That different
angular distribution yields a different far-field radiation pattern
emerging from the fiber into the spectrograph.  This effect is known
as geometrical focal ratio degradation (FRD).  The emerging radiation
pattern will propagate through the spectrograph optics to yield a
different PSF (e.g., the PSF may have a tilt-dependent centroid shift
in the wavelength direction).

\item Focal ratio degradation (FRD) in the fibers: light exits an
  optical fiber with a slightly wider angular distribution (a faster
  beam) relative to the fiber axis than it had upon entering the
  fiber.  This form of FRD depends on the stress on the fiber due to
  its routing from the focal plane to the spectrograph.  In MSE, FRD
  could decrease the focal ratio capturing 95\% of the light by
  $3 - 6$\%, placing some light at a given wavelength outside of the
  acceptance speeds of the collimator and altering the PSF of the
  accepted light.  The effect of a FRD-dependent PSF on the sky lines
  is in principle removable through a principal component
  analysis. 

FRD effects can also be minimised at the design
  stage of the SIP.FiTS subsystem. Fibers leading to the high
  resolution spectrograph are approximately 50m long, and fibers
  leading to the low/moderate resolution spectrographs are
  approximately 30m long. Careful fiber-routing and
  strain-relief are therefore essential to reduce any twisting effects or
  any other action that stresses the fibers. 

\item Spectrograph optics: MSE requirements impose a minimum
  resolution and an acceptable range for the average resolution.
  However, no requirement is placed on resolution variations at the
  same wavelength for fibers imaged by the spectrographs at different
  locations on the detectors.  Such variations have been reported in
  the conceptual design report for the low/moderate resolution
  spectrograph.  We note that such variations in PSF spectral
  {\it width} can be accounted for more easily than variations in PSF
  spectral {\it shape}.

\item Detector effects (including pixel-to-pixel sensitivity variations, and charge diffusion):
To achieve continuum sky subtraction accurate to 0.5\% at any
wavelength, pixel-to-pixel sensitivity variations on the detectors
will have to be calibrated so as to contribute $<0.5\%$ uncertainty to
the sky per extracted 1-D pixel.  For example, if a spectral trace is
three pixels across in the spatial direction, the response of each
pixel must be calibrated to $<0.87\%$ each to yield a combined
$<0.5\%$ in the corresponding 1-D spectral pixel.

\item Fiber-to-fiber wavelength calibration errors.  The wavelength
  calibration will depend on fitting numerous individual lines. Shifts
  in the wavelength centroid due to fiber tilt affecting the PSF are
  estimated to be very small.

\item Spectrograph thermal or mechanical changes.  Neither the LMR or
  HR spectrographs will be located on the telescope elevation
  structure, so spectrograph flexure during exposures is not an
  issue. However, mechanical stability in the LMR spectrograph due to
  grating changes between the LR and MR modes is a potential issue.
\end{enumerate}

Also required is knowledge of the properties of any significant
spectrograph stray light or ghost image across each detector in each
spectrograph. Such light could mimic sky background light but with
different wavelength and spatial variations; such light would have to
be modeled and removed.  

\subsection{Considerations for spectrophotometry}

Absolute spectrophotometry is the conversion of recorded photon counts
at each wavelength in a spectrum to an object’s total flux density
(traditionally in ergs per cm$^2$ per Angstrom).  Relative
spectrophotometry is the reconstruction of an object’s relative flux
density at each wavelength, yielding accurate spectral shapes but not
photometry (i.e., with an unknown normalization factor).  Relative
spectrophotometry is desired for low and medium resolution spectra.
For high-resolution spectra, there is no spectrophotometry
requirement, but calibration stars can be observed when a correction
for telluric absorption.

Flatfield calibrations (pixel flats and spectral flats) provide the
relative throughput as a function of wavelength for all fibers in a
given observing setup, enabling conversion from recorded counts
(photoelectrons) in a fiber to recorded counts normalized relative to
a reference fiber (or fibers, or other reference value).  

However, flatfield calibrations do not correct for the injection
efficiency (IE). The IE is the fraction of light from an astrophysical
object incident at the focal plane that enters the fiber, and it
depends on the distribution of object light at the focal plane and the
relative positioning accuracy of the fibers. The former depends on all
terms that contribute to the image quality - including the
polychromatic seeing of the free atmosphere, thermal and airflow
effects in the enclosure (dome seeing), the residual effects of
atmospheric dispersion not corrected for by the ADC, atmospheric
refraction, and all optical effects in the M1 system and WFC/ADC - and
is potentially able to be modeled given excellent knowledge of the
system. The latter depends upon numerous hardware performance issues,
especially the fiber positioners but other effects as well. IE is
discussed at length in Ref~\citenum{flagey2018c}.

Converting flatfielded counts to relative flux densities
requires observing targets with a known spectral energy distribution
(SED. In practice, for a fiber-fed multi object spectrograph, this
requires observing many ($\sim20$ or more) relatively bright, hot stars
of known, constant magnitude whose spectra can be modeled to high
accuracy.  Such spectrophotometric calibration stars are spread over
the field of view and are observed simultaneously with science
targets. Each model spectrum, normalized to the star’s known
magnitude, is divided by the counts at each wavelength to compute a
``fluxing vector'' i.e., the transformation between the flatfielded
calibration star counts and the physical SED.  These fluxing vectors
are typically averaged to produce a final fluxing vector which is
multiplied by the flatfielded counts of all targets in the exposure to
yield final flux-calibrated spectra.

In the SDSS-I/II survey, which used 3'' diameter fibers and no ADC, an
RMS spectrophotometric accuracy for calibration stars of 4\% at a
given wavelength was achieved on average.  In the SDSS-III/BOSS
survey, which used 2'' diameter fibers and the same site and telescope
as SDSS-I/II, the corresponding number was 6\%.  For comparison with
the MSE relative spectrophotometry requirement, note that the
uncertainties on the $g-r$ and $r-i$ colors [equivalent to flux
ratios] in BOSS were 5.7\% and 3.2\%, respectively, or 4.5\% on
average.  In the SDSS-RM campaign (part of SDSS-III), the
corresponding number was 5\%.  SDSS-RM observed its targets twice as
long as normal SDSS-III targets and was thus somewhat more susceptible
to the effects of atmospheric dispersion, but was able to compensate
for that effect and even improve the calibration slightly by using 70
spectrophotometric standards instead of 20.

Maximizing the number of fibers on science targets makes it practical
to observe only enough calibration stars to calculate a single fluxing
vector per exposure.  Therefore, known systematic effects on the
wavelength-dependent injection efficiency of different fibers across
the field of view should be accounted for by removing those effects
before calculation of the fluxing vectors. If all known systematic
effects have been removed, then the scatter in the fluxing vectors
will be minimized and the accuracy of the spectrophotometry maximized.
It is worth emphasizing that effects which vary across the field of
view must be corrected before calculation of the fluxing vector, but
that effects which are uniform across the field of view can be
incorporated in the fluxing vector.

If only random uncertainties remain to limit the spectrophotometric
accuracy, then the spectral shape of a target will be as accurately
characterized as possible even when it is observed repeatedly with the
telescope at different elevations and azimuths, using different fibers
at different locations in the focal plane, at different fiber tilts,
and recorded on different detectors in different spectrographs.

Finally, we note that the approach used to meet MSE's relative
spectrophotometry requirement will also yield absolute
spectrophotometry of comparable accuracy for point sources.  Our first
step toward relative spectrophotometry is a correction for flux not
intercepted by each fiber as a function of wavelength.  We only
require that step to be accurate in a relative sense.  In the next
step, the fluxing vector incorporates whatever correction factor is
needed to match the absolute fluxes of the spectrophotometric
calibration stars, with an accuracy determined by the known photometry
and the modeled spectral shape of those stars.  The result is
spectrophotometry which is accurate in both a relative and an absolute
sense, though we do not consider here any additional uniformity
considerations relevant to absolute spectrophotometry.

\section{OPERATIONAL CONSIDERATIONS}\label{obs}

We now discuss (nighttime) operational considerations that
impact the calibration of the data.

\subsection{Observing efficiency requirement and definition}

MSE is a survey facility, whose success will be primarily related to
the quantity and quality of data obtained every night. There is a high
level science requirement that states that the ``observing
efficiency'' of MSE will be 80\%. Specifically, observing efficiency
is defined as the fraction of time the observatory is collecting
photons divided by the time the observatory could have been collecting
photons, which is all the time available for observations except that
lost to weather. We refer the reader to Ref.~\citenum{flagey2018b} for a
detailed discussion of observing efficiency in the context of MSE.

The observing efficiency is defined in “steady state” operations for
MSE, i.e. after commissioning of the observatory. In addition, we
assume it is averaged at least over a year, given the nature of the
typical events occurring at a ground based astronomical facility. 

The intent of this requirement is to ensure the efficient acquisition
of quality data. Some calibration observations will be required during
the nighttime, and while these officially count ``against'' the
observing efficiency, it is not deemed acceptable by the Project to have a high
acquisition rate of data that cannot be used for science. The
challenge is therefore to minimise nighttime calibration time, while
ensuring an efficient acquisition rate of science-quality data.

\subsection{On-sky calibration time}

The ``worst-case scenario'' for nighttime lamp calibrations with MSE is
that they are expected before and after each science observation
(which we term an ``Observing Matrix'', OM, in what follows). In order
to mesh with other aspects of night-time operations and the overall
observing efficiency budget, we aim for a nighttime calibration
sequence to last no more than 4 minutes, broken down as two blocks. We
assign the timing of these blocks as follows:

\begin{itemize}
\item 95 seconds for collecting photons (TBC)
\item 18 seconds for all readouts (TBC)
\item 3.5 seconds to turn on the system (TBC)
\item 3.5 seconds to turn off the system (TBC)
\end{itemize}

The justification for this break-down is as follows:

\begin{itemize}
\item A nighttime calibration sequence will use the exact same configuration
of the telescope as during the corresponding OM. No additional time
will be required to configure the system, apart from turning on the
calibration unit. The ``calibration time'' is the sum of the time
spent collecting photons, reading out the detectors, and turning the
calibration system on/off.

\item There will be two different sets of calibration exposures to obtain at
night: flat and arcs. There will possibly also be a set of
calibrations for low/moderate resolution, and another for high
resolution, given that both modes operate continuously. We baseline
multiple exposures for both flats and arcs, to mitigate issues that
could occur on a single exposure (e.g. cosmic rays), although we
ultimately aim to only require a single exposure. With a baseline of 3
arcs and 3 flats for each set of calibrations, we need to allocate
time for a total of 12 calibration exposures.

\item Detectors read-out with low noise (a few electrons) can occur at a
frequency of about 1 MHz (e2v 231 series, 6k by 6k, 3 MHz max, 5e- at
1 MHz, 2e- at 50 kHz). For calibration exposures, low read-out noise
is not be necessary and the fastest read-out rate will be used (12
seconds). Binning (2x2 for HR, 2x1 for LMR) will shorten the readout
time to 6 seconds, and using all 4 outputs will decrease it to 1.5
second. The total allocated time for all readouts is thus 18 seconds.

\item We allocate 7 seconds to switching the system on/off (3.5 seconds
each). This includes moving any mechanical part of the calibration
system (e.g. deploying a screen). Some of this time will be spent in
parallel with other processes.

\item Each nighttime calibration block is limited to a reasonable
allocation (2 minutes). We therefore have 95 seconds left to allocate
to the time spent collecting calibration photons.
\end{itemize}

\subsection{Night-time observing sequence}

   \begin{figure} 
   \begin{center}
   \includegraphics[width=15cm]{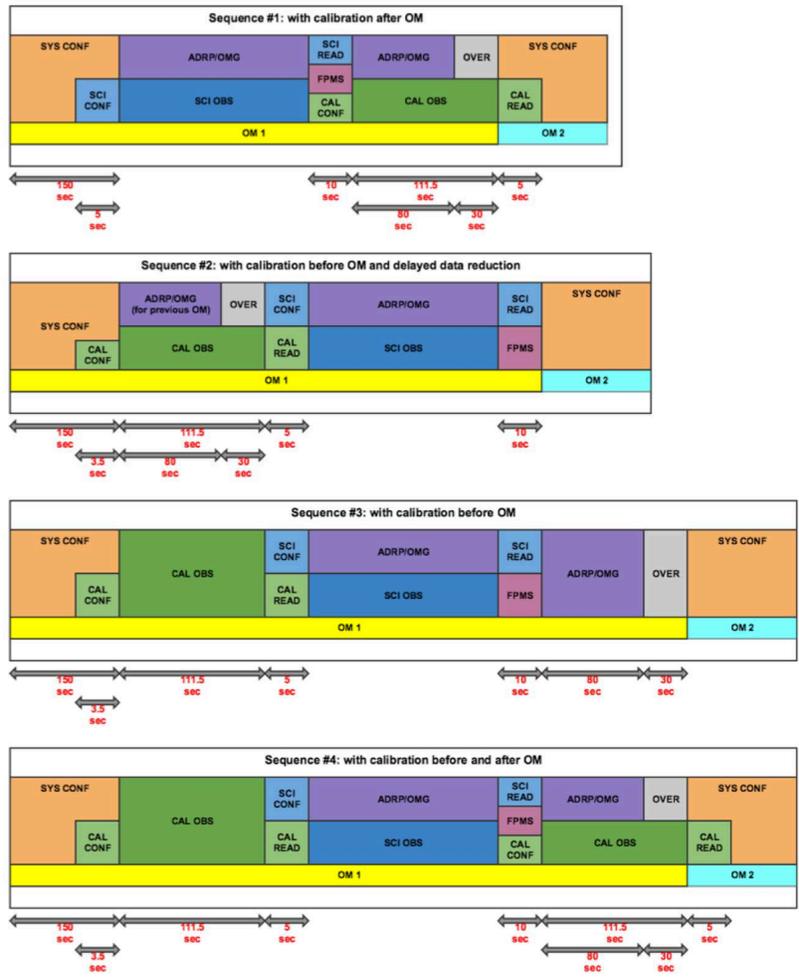}
   \end{center}
   \caption[example] 
   { \label{fig:seq} 
Typical nighttime sequences, baseline at bottom (calibration exposures
both before and after the science exposure). See text for details.}
   \end{figure} 

Given the time spent on calibrations, we now illustrate how we can
reconcile our calibration needs with the observing efficiency
requirements. We assume a typical science observation of 1 hour in
what follows.

Figure~\ref{fig:seq} shows typical nighttime sequences, with the following
definitions for the blocks:

\begin{itemize}
\item SYS CONF: system configuration (telescope, enclosure, positioners, …)
\item SCI OBS: science observations (collecting photons from science targets, not including last readout)
\item SCI READ: science readout (last readout of the SCI OBS)
\item FPMS: measuring positions of fibers after SCI OBS
\item CAL CONF: calibration configuration (turning on calibration unit)
\item CAL OBS: calibration observations (collecting photons from calibration unit, not including turning it on/off and last readout)
\item CAL READ: calibration readout (last readout of the CAL OBS, turning off the calibration unit)
\item SCI CONF: science configuration (usually after a CAL OBS to make sure the guiding is still good for science observations)
\item ADRP/OMG: automatic data reduction pipeline and updating schedule
\item OVER: MSE Staff override of the schedule update
\end{itemize}

In all sequences, we assume the following processes can occur in
parallel:

\begin{itemize}
\item SCI CONF, SYS CONF, CAL CONF: the elements in SCI CONF (guiding
  check) and CAL CONF (turning calibration unit on) are part of SYS
  CONF when this block is followed by a SCI OBS and CAL OBS,
  respectively.
\item SCI READ, CAL CONF, FPMS: while the last science exposure is
  readout, the calibration system is turned on, and the FPMS
  accurately measures the positions of the fibers at the end of the
  OM.
\item CAL OBS, ADRP/OMG, OVER: while the calibration exposures are
  being obtained and readout, the real-time feedback to the scheduler as
  well as the scheduling will occur. In addition, we allocate some
  time for the MSE Staff to override the decision of the scheduler.
\item SYS CONF, CAL READ: while the last calibration exposure is
  readout and the calibration system is turned off, the system will
  configure for the next OM.
\item SCI CONF, CAL READ: while the last calibration exposure is
  readout and the calibration system is turned off, the system will
  verify guiding is still active.
\end{itemize}

The sequences in Figure~\ref{fig:seq} differ by the order in which the CAL OBS and
ADRP/OMG are executed. The baseline for MSE is to use the sequence at
the bottom (Sequence 4) with calibration exposures obtained right before
and right after the OM, for improved data reduction and
calibration. The other three
sequences each only obtain a single calibration observation (before or
after the science observation); other differences between these
sequences relate to when certain events occur relative to others, and
is outside the immediate scope of this paper.

Figure~\ref{fig:seq}  indicates the time allocated to each block and we use it to
derive the total amount of overhead per OM in each sequence. For the
baseline sequence (Sequence 4), overheads sum to 388 seconds. Given an
average OM duration of 60 minutes, these overheads lead to a total
time per OM of 3988 seconds.

Given the average time available per night, we then derive the average
number of OM per night. Combining with other elements of the Observing
Efficiency calculation, such as mean failure rates, engineering time,
etc, leads to an estimated observing efficency of 82\%. Details can be
found in Ref.~\citenum{flagey2018b}. Thus, even
with calibration exposures occuring before and after each science
exposure, we still expect to meet the (demanding) observing efficiency
requirement.

\section{OVERVIEW OF CALIBRATION PROCEDURE}

This section describes the entire suite of observations - both daytime
and nighttime calibrations, lamps and contemporaneous targets - that
are envisioned as being required to provide suitable calibration for
MSE.

\subsection{Biases and Darks}

Biases should be obtained at the start and/or end of every night, and
scripts set up to take bias frames when the telescope is not otherwise
in use. Ideal bias frames are the median of a very large number of
bias frames all taken close in time to the science observations to
which they are applied. This follows “best-practice” techniques for
calibration exposures, but we clearly hope that the bias frames do not
change significantly with time.

Dark frames are, ideally, not necessary, since modern optical
detectors generally do not have a large dark current. However, they
may be necessary for NIR channels. If they are required for either the
optical and/or the NIR they should be obtained as part of a daytime
procedure (likely taken the day after the nighttime science, in order
to match the actual exposure times used). 

Even in the case of dark frames proving to be unnecessary, it is good
practice to take occasional dark frames (weekly or monthly) to monitor
the detectors and ensure that their characteristics have not changed
unexpectedly (which would indicate an issue for trouble-shooting).

\subsection{Continuum flats}
 
\subsubsection{Lamp flats}

The adopted baseline for MSE for flat-fields will require a sequence
of high SNR “lamp” flats to be observed immediately before and
immediately after the science exposure. This will ensure that the
flats are obtained with the entire system in the same setup as used
during the science exposure, in environmental conditions as close as
possible to the science exposure. 

It is essential that these lamp flats illuminate the focal surface
with a repeatable illumination pattern and Spectral Energy
Distribution which is the same for every fiber, at a level not
compromising either spectrophotometric or sky-subtraction-precision
requirements.

It is very desirable that the far-field light arriving into the
spectrographs from these lamp flats reproduce the beam arriving from
the sky as closely as possible. This is because small errors in the
modeled spatial PSF translate into significant errors in correcting
for cross-talk.

Obtaining calibration exposures at the beginning and end of the
science exposures ensures that any differential changes in the system
and/or environment occurring during the science exposure (e.g., due to
the InRo) will be captured by the calibration exposures.

Since these exposures are being taken during the night, appropriate
hardware is required in order to ensure the observations can be done
quickly and so not impact the observing efficiency of MSE more than is
already discussed in the previous section. We discuss this possible
hardware configuration in Section~\ref{SCal}.

\subsubsection{Twilight flats}

The adopted baseline for MSE for flat fields will additionally require
obtaining as many twilight flats at the beginning and end of the night
as is possible. These flats are used specifically to ensure that all
fibers have an even illumination to a high level. This will be
accomplished by taking the median of numerous twilight flats, and if
needed correcting for intrinsic non-uniformity in the twilight sky
across the field.

Lamp flats, or even dome flats, do not provide the level of uniformity
that is possible via a twilight flat.

\subsubsection{Combining twilight and lamp flats to obtain the
  transfer function}

At the beginning of the night, and again at the end of the night, a
sequence of twilight flats are obtained with the entire MSE system
(including telescope, PFHS, InRo, fiber positioners, etc.) in some
reference position. The median of these flats provides a reference
frame with even illumination across all fibers (hereafter, the “master
twilight flat”).

Immediately at the end of twilight, a series of lamp flats are taken
with the telescope and system in exactly the same configuration as for
the twilight flats. The median of these frames (hereafter, the
“reference lamp flat”) provides a suitable reference to connect the
lamp flats to the twilight flats.

At the beginning and end of every science exposure, a series of lamp
flats are taken with the telescope and system in the same
configuration as will be used (or, will have been used) for the
science exposures. For long exposure sequences in which fiber
allocations do not change, the option will be available to reduce the
number of lamp flats taken during the sequence. including taking just
one lamp flat in between exposures, instead of two. Each individual
calibration lamp flat exposure obtained in this way is hereafter
referred to as a “science lamp flat”.

The “Master Science” flat field, that defines the transfer function to
be applied to the data, is then given by:

\begin{equation}
Master~Science~Flat = Median \left[ science~lamp~flat \times \left( \frac{master~twilight~flat}{reference~lamp~flat} \right) \right]
\end{equation}

Based on experience with other fiber spectrographs, we do not
anticipate using dome flats as part of the standard MSE calibrations.
However, a flat-field screen will exist for use with dome arcs, and so
the possibility exists of designing an illumination system to provide
dome flats.

\subsection{Arc calibration exposures}

\subsubsection{Lamp arcs}

Lamp arcs are obtained in exactly the same fashion as lamp
flats. Specifically, the adopted baseline for MSE for arc calibration
will require a sequence of high SNR “lamp” arcs to be observed
immediately before and immediately after the science exposure. There
is no requirement for a uniform intensity of arc illumination across
the focal surface, provided that the signal is strong enough
everywhere that good wavelength solutions can be derived for all
fibers. However, It is essential that both the near-field and
far-field light arriving into the spectrographs from these lamp arcs
reproduce the light arriving from the sky, at a level that does not
compromise the required velocity precision or sky-subtraction
precision (via errors in knowledge of the PSF). All other
considerations are the same as for lamp flats.

\subsubsection{Dome arcs}

The adopted baseline for MSE for arc calibration additionally requires
sequences of high SNR dome arcs. We baseline these as daytime
observations to be taken daily. Dome arcs are required in order to
illuminate the system with arc light with a similar far field pattern
as the science light, which will affect the PSF. Lamp arcs may not
fulfill this requirement, and comparison of lamp arcs to dome arcs
will provide the information necessary to make any required
corrections to the lamp arcs. 

\subsubsection{Combining dome and lamp arcs to obtain precise
  wavelength calibration}

During the day, a sequence of dome arcs will be obtained with the
entire MSE system (including telescope, PFHS, InRo, fiber positioners,
etc.) in some reference position. The median of these arcs provides a
reference frame with a far field illumination that matches the science
observations (hereafter, the “master dome arc”).

Immediately before and after these dome arcs, a sequence of lamp arcs
are taken with the telescope and system in exactly the same
configuration as for the dome arcs. The median of these frames
(hereafter, the “reference lamp arc”) provides a suitable reference to
connect lamp arcs to dome arcs.

At the beginning and end of every science exposure, a series of lamp
arcs are taken with the telescope and system in the same configuration
as will be used (or, will have been used) for the science
exposures. Each individual calibration lamp arc exposure obtained in
this way is hereafter referred to as a “science lamp arc”.

Different far field illumination patterns can yield different
wavelength solutions (via different line spread functions, for
example).  If the dome arcs and the science exposures have identical
far field illumination, then dome arcs taken in the same configuration
as the science lamp arcs would provide the most accurate wavelength
solution.  Because such arcs will not be available for MSE, the
following approach is used to account for the possibility of
wavelength solution differences between lamp arcs and dome arcs, and
between lamp arcs obtained in science configurations and in the
reference configuration in which dome arcs are obtained.

The science lamp arc wavelength solution is compared to the reference
lamp arc wavelength solution to derive a transformation between
solutions in the different system configurations. The final adopted
wavelength solution is the master dome arc wavelength solution
corrected by this transformation.  For example, in a simple case in
which different PSFs produce wavelength solutions which differ only in
their zeropoints, the adopted wavelength solution is the master dome
arc wavelength solution corrected by the zeropoint offset between the
reference lamp arc and science lamp arc wavelength solutions.)

Ideally, dome arcs should use the same sources as the lamp arcs, and
certainly the same type of sources, to ensure there are no additional,
unnecessary, systematic differences between the two sets of arcs.

\subsection{Additional considerations regarding procedures for
  calibration lamps}

Since both high resolution and low resolution spectrographs are used
simultaneously by MSE, the lamp calibrations must be useful for both
modes. It is a goal that this can be obtained in a single set of flats
and a single set of arcs, but we note it is possible that the lamp arc
calibration procedures will need to be repeated for both the low
resolution and high resolution modes, and this is what is currently
budgeted in the observing efficiency budget described in Section~\ref{obs}.

As more data is obtained by MSE, and as we obtain more and more
calibration exposures, new opportunities might become available to
improve the efficiency of lamp calibration observations (flats and arcs):

\begin{itemize}
\item We should be able to determine the feasibility or otherwise of
  obtaining calibrations at the start or end of science exposures,
  instead of both at the start and at the end;
\item Comparisons of twilight flats and master twilights taken as a
  function of time should reveal the need, or otherwise, to obtain
  these at the start and end of every night (potentially, the
  frequency with which they are obtained could decrease);
\item Similarly, comparison of reference lamp flats with time will
  reveal the frequency with which these need to be obtained;
\item Comparison of nighttime science lamp calibrations (flats and
  arcs) between exposures taken as a function of system set-up
  (including InRo position, fiber positions, time, etc.) should reveal
  the need or otherwise to obtain science lamp flats for every
  individual science set up, versus some other frequency based on the
  observed behaviour;
\item The frequency of dome arcs may be able to be decreased, if it is
  clear that the daily dome arcs do not vary significantly with time.
\end{itemize}

Additionally, during the day, some calibration exposures (lamp flats
and arcs) should be repeated. That is, the system
should repeat the sequence of moves that it went through the previous
night, and lamp exposures should be repeated as they were during the
night:

\begin{itemize}
\item Over time, the comparison of daytime lamp calibration exposures
  versus the corresponding night time exposures will inform us on
  whether there is a way to utilize daytime science lamp exposures in
  place of some or all nighttime science lamp exposures;
\item It is a goal of MSE to be able to reduce or remove the need for
  nighttime science lamp exposures to improve observing efficiency, so
  long as the science utility of the calibrated science data are not
  affected detrimentally.
\end{itemize}

\subsection{Pixel flats}

The detector introduces an additional dependency of throughput,
response (etc.) on the pixels (spectral flats and arcs measure the
system response as a function of wavelength, including but not limited
to the detector). Thus, to remove this pixel dependency requires
creating a flat field that measures primarily the response on the
detector, not the fibers or the rest of the system.

A direct way of making this measurement is to put a lamp inside the
spectrograph, that illuminates the collimator, and hence detectors,
uniformly. It can prove tricky to get the desired level of
uniformity. For SDSS, pixel flats are taken rarely (every 6 months or
so), since the procedure involves changing the entire slit-head with a
“leaky fiber”, that evenly illuminates the slit.

An alternative to putting lamps inside the spectrographs is to create
a pseudo-pixel flat by taking a large number of (possibly defocused)
spectral flats (the defocus might be necessary to put light between
the spectral traces). These flats would be median-combined. The
resulting frame would then be divided by a 2D model of itself (i.e. a
smooth spectrum multiplied by the fitted spatial profiles). This would
then be divided by a locally median-filtered version of itself, to
remove residuals from the imperfect 2D model and give a pixel flat. In
principal defocusing is not needed, since the raw frames are much
brighter than the data frames in every pixel, and there are many of
them. This procedure is an elaboration on that successfully used for
AAOmega, which adopted this process after unsuccessful use of lamps
inside the spectrographs, and is the adopted baseline for MSE.

\subsection{Diffuse light}
The diffuse light background from spectrograph stray light must be
modeled at a level so as to not compromise the spectrophotometric and
sky-subtraction accuracy.  Sufficient space at the edges of the
detectors and gaps between selected spectra on the detectors must be
allocated to measure the spectrograph stray light to a precision
better than 1\% of the sky flux in neighboring spectra and to fit a
plausible model to the two-dimensional distribution of such light on
the detector.  Such edge and gap locations are needed so that the
contribution of the wings of the PSF from the neighboring spectra is
minimized. While the wings of the PSF are a type of “diffuse light” by
some definitions, they are a very local source and our intent here is
to measure the overall background due to spectrograph and telescope
structure stray light.

\subsection{Contemporaneous observations: sky spectra and
  spectrophotometric calibration stars}

Spectrophotometric calibration stars will need to be observed
contemporaneously to science targets. The number of these per field
will be defined primarily via the spectrophotometry requirements, and
further analysis is required to determine this number. We note that,
given the expected limited number, spectrophotometric calibration
stars will likely need to be assigned in a given field with the
highest priorities (i.e., before most science fibers have been
assigned to fibers). Ideally, these will be distributed across the
field and will be distributed between the banks of spectrographs.

The number of fibers assigned to sky is expected to be around 10\%,
and these should likely be distributed evenly across the field and
will be distributed between the banks of spectrographs. It is expected
that subject to a minimum sampling of the slit for every spectrograph,
the sky fibers can be assigned with the lowest priority; that is, that
any fibers not able to be allocated to science targets will instead be
allocated to sky.

\section{SCIENCE CALIBRATION HARDWARE REQUIREMENTS}\label{SCal}

\subsection{General considerations}

Consideration of the calibration procedure described above suggests
the following:

\begin{itemize}
\item Lamp flats/arcs must provide enough photons, for all modes,
  fibers and wavelengths in a matter of seconds, allowing them be
  obtained at a high SNR in a reasonable amount of time. Since this
  must be done without saturating the detectors (except perhaps for a
  known subset of arc lines), there are stringent constraints on the
  allowed variation in flux with wavelength or spectral
  line. Calibration read-out times need to be as short as
  possible. Ideally, fast read-out options should be available, and
  the SNR of the calibration exposures will be high enough such that
  these can be used without affecting the overall SNR of the
  exposures;
\item To minimize overheads, solutions that close the dome or deploy a
  screen onto which the lamps are shone during the night, are not
  feasible for standard operations; 
\item As the concept below is developed further, we need to
  investigate the consequences of having lamps illuminated in the dome with the
  shutter open for the other telescopes on the mountain. If the lamps
  are ``faint'', then this shouldn't be an issue but will be seriously investigated;
\item We do not want to spend a significant time per nighttime
  calibration observation waiting for the lamps to switch on and warm
  up. Either the lamps will have to warm up and stabilize in less than
  the science exposure readout time, or they must be used with a
  shutter, and be capable of continuous use without compromising heat
  or light leakage requirements;
\item During the night, the dome aperture will be aligned with the
  telescope. Therefore, when calibration exposures are being executed,
  photons from the sky will also be collected (as well as the
  targets). All arc-line or continuum fitting, flat-fielding, etc.,
  will have to allow for this effect. Several options exist to remove
  this effect:
\begin{itemize}
\item The severity of the effect will be reduced with reduced exposure
  times (i.e., brighter calibration lamps are preferred);
\item Calibration arc or flat exposures can be accompanied by a
  ``calibration sky'' exposure, ideally of identical exposure time and
  with a fast
  readout time. The ``calibration sky'' exposure would be subtracted
  from the calibration exposures themselves, after scaling for
  exposure time differences if needed, to remove the object and sky
  flux from each fiber.
\item We could scale the data frames according to time (and change in
  gain) and subtract them from the calibration exposures, potentially
  removing both the object and sky signatures simultaneously;
\item We also considered offsetting the telescope a few arcsecs in
  azimuth, to avoid changes in airmass or gravity vectors, and to
  place the fibers on empty sky, particularly for calibration
  exposures at the end of a science exposure. Guiding would continue
  with the same guide stars, by switching the guide stars between two
  different sub-rasters on the guide cameras.  Guiding contributes
  only 2 microns RMS to the fiber positioning uncertainty; the
  uncertainty in returning to the same position after offsetting away
  and back would be of the same magnitude. However, calibration
  exposures taken at this position still contain sky light, and so
  this is an incomplete solution.
\end{itemize}
\end{itemize}

\subsection{Lamp flats and arcs}

The baseline calibration lamp concept for MSE has the calibration
lamps (flats, arcs, appropriate for all spectral resolution settings
of MSE) distributed along the underside of each strut of the telescope
structure (or fiber outputs distributed along the underside of each
strut, where the fibers are fed by the appropriate calibration lamps)
and which point down onto M1. The distribution of the lights along the
struts of the telescope will be such that the radial light
distribution mimics that of the telescope as closely as possible. When
these lamps are switched on, the light is incident on M1 and sent to
the focal plane of the telescope. There, we rely on the azimuthal
scrambling properties of the fibers to ensure that the azimuthal light
output from the fibers has a uniform intensity and does not possess a
memory of the initial light distribution on the struts. 

We note that the distribution of lights (either in radius or in
function, e.g., arcs, flats) can be staged for each strut, since what
matters is the overall radial distribution.

It is TBD whether fixed sources on the underside of the struts will be
sufficient, or whether these sources need to be on movable
stages. Movable stages will almost certainly produce a smoother light
distribution than is possible with discrete, fixed, locations, but
will increase the complexity of the system significantly. However, it
is expected to be more likely that the lamps at largest radius have to
be on movable stages, to avoid discrete jumps in the illumination of
the focal surface due to WFC vignetting.

It is TBD whether we can use the limited radial scrambling properties
of fibers to help “smooth” the radial distribution of light, to ease
the issue of discretization due to multiple sources.

The precise nature of the lamps is open, as is the question as to
whether single sets of lamps will suffice, or if different
configurations will be needed for different arms or resolutions. The
lamp flats would ideally provide useable photon fluxes at all
wavelengths and spectral resolutions simultaneously (albeit with
curtailed exposure times in some arms/resolutions). The lamp arcs must
provide useable densities and fluxes of spectral lines at all
wavelengths and resolutions, preferably simultaneously, and again with
curtailed exposure times as needed.

\subsection{Dome arcs}

Ideally, the dome arcs should use the same sources (and certainly the
same type of sources) as the lamp arcs to ease all comparisons between
dome arcs and lamp arcs. 

Hollow cathode arc lamps are intrinsically very faint, and so a large
number of lamps and/or long exposures would likely be required to
obtain arcs with sufficient SNR to be useful. 

There must be an area on the interior of the dome that the telescope
can point to for science observations, that is white and to which the
telescope can point to obtain dome arcs. This could take the form of a
deployable or fixed screen.

\section{CONCLUSIONS AND FUTURE WORK}

In this paper, we have discussed the operational and subsystem issues
pertaining to accurate scientific calibration of science spectra
obtained with MSE. The operational strategies and proposed
observations are designed to provide the user with sufficient
empirical information on the performance of MSE to enable estimation
of the differential throughput, PSF behavior and wavelength
solutions for all the objects observed in each observation. This
information is necessary to allow accurate sky subtraction, spectrophotometry and
velocity estimation. We note that we have not discussed the analysis methods by
which we use this information to do precise sky subtraction; this is
an extensive subject by itself that will be discussed in future
contributions.

MSE will shortly undertake a conceptual design of the SCal subsystem
during which the detailed requirements of this unit and the
calibration strategy in general will be refined. In addition, MSE has
formed a Calibration Working Group. This group consists of three
people from Europe, the US and Australia that have extensive practical
experience in wide field fiber MOS. They will advise the Project on
all matters relating to calibration, including reviewing all relevant
material and offering solicited and unsolicited feedback on any
aspects of MSE that they deem relevant to the goal of obtaining high
quality calibrated science data. Finally, MSE is starting to develop a
Design Reference Survey, that is a detailed simulation of an actual 2
year observing program conducted with MSE, and which will include
calibration observations, both contemporaneous on-sky calibrators and
nighttime lamp calibration exposures. Updates on all aspects of these
developments will be provided in future SPIE contributions.

\acknowledgments 
 

\end{document}